# Reproducibility of Implicit Association Test (IAT) – Case study of meta-analysis of racial bias research claims

S. Stanley Young[1] and Warren B. Kindzierski[2]

[1] CGStat, Raleigh, NC, USA
[2] Independent consultant, St Albert, Alberta, Canada
Correspondence: Warren B. Kindzierski, Email: warrenk@ualberta.ca.

**Abstract**

The Implicit Association Test, IAT, is widely used to measure hidden (subconscious) human biases – implicit bias – of many topics: race, gender, age, ethnicity, religion stereotypes. There is a need to understand the reliability of these measures as they are being used in many decisions in society today. A case study was undertaken to independently test the reliability of (ability to reproduce) racial bias research claims of Black−White relations based on IAT (implicit bias) and explicit bias measurements using statistical p-value plots. These claims were for IAT−real-world behavior correlations and explicit bias−real-world behavior correlations of Black−White relations.

The p-value plots were constructed using data sets from published literature and the plots exhibited considerable randomness for all correlations examined. This randomness supports a lack of correlation between IAT (implicit bias) and explicit bias measurements with real-world behaviors of Whites towards Blacks. These findings were for microbehaviors (measures of nonverbal and subtle verbal behavior) and person perception judgments (explicit judgments about others). Findings of the p-value plots were consistent with the case study research claim that the IAT provides little insight into who will discriminate against whom. It was also observed that the amount of real-world variance explained by the IAT and explicit bias measurements was small, less than 5%. Others have noted that the poor performance of both the IAT and explicit bias measurements are mostly consistent with a 'flawed instruments explanation' – problems in theories that motivated development and use of these instruments.

# 1. Introduction

*1.1 Implicit Association Test*

The Implicit Association Test (IAT) is a tool developed by psychologists (Greenwald et al. 1998). It is a visual and speed reaction test done on a computer where a subject associates words and pictures. The test claims to measure unconscious bias towards a topic of interest in our population today – e.g., race, gender, social status, etc. The test speed of association is taken to measure the 'strength of association' between pictures and words, and, by assertion, measures unconscious bias, say of Whites towards Blacks (racial discrimination), males towards females (sex discrimination), wealthy people towards poor people (wealth discrimination), etc. For example, the majority of White Americans who have taken the IAT have been classified as anti-Black (Blanton et al. 2009). There are other established tools for measuring such biases, e.g., explicit questionnaires (e.g., Wittenbrink et al. 1997) and observations of real-world behaviors (e.g., Oswald et al. 2013a).

The IAT is widely used by academic researchers in the fields of psychology, sociology, neuroscience, and social sciences. A *Google Scholar* search of the phrase "Implicit Association Test" on 24 November 2023 returned over 24,000 articles and/or citations. It is popular to look at disparities (differences) in discriminatory behaviors using the IAT in diverse practices such as healthcare (Maina et al. 2018), education (Harrison & Lakin 2018), employment and hiring (Ziegert & Hanges 2005), criminal justice (Bass 2021). It is noted that implicit bias is just one of many other factors that may explain some disparities in these practices (e.g., age, experience, etc.).

The IAT is also used in business as a tool for dealing with DEI (diversity, equity, and inclusion) issues in the workplace; for example, Rand Corporation (2023). It is also popular within universities (e.g., University of Ohio – Sailer 2023). Randall (2023) recently described DEI training being imposed by the US government via presidential executive order in: the Centers for Disease Control and Prevention, the Department of Education, the Department of Justice, the Office of Science and Technology Policy, and the Office of Personnel Management.

*1.2 Validity and Reliability*

The IAT is not without its problems. In the past others have criticized the validity (accuracy) and reliability (reproducibility) of the IAT; for example: Fiedler et al. (2006), Blanton et al. (2009), Oswald et al. (2013a and 2015), Schimmack (2019 and 2021), Mitchell & Tetlock (2020). Psychometricians – scientists who study measurements of people's knowledge, skills, and abilities – question the validity of the IAT "construct", i.e., the underlying validity of an IAT measurement. More specifically, does an IAT measurement actually measure what it purports to measure?

To establish the validity of an IAT measurement, one can look at relationships between that measurement and other measurements. These measurements should be correlated with each other. In these terms, IAT measures of unconscious bias (implicit bias) should be correlated with explicit measures and observations of real-world behaviors or actions displaying bias.

The objective of our case study was to independently test the reliability of (ability to reproduce) racial bias research claims of Black−White relations, specifically negative behaviors of Whites towards Blacks, based on IAT and explicit measurements. It involved using statistical p-value plots (Schweder & Spjøtvoll, 1982) and publicly available data sets to visually inspect the reproducibility of meta-analysis cause−effect research claims. These claims were for IAT−real-world behavior correlations and explicit bias−real-world behavior correlations of behaviors of Whites towards Blacks.

## 2. Methods

We initially developed and posted a research plan for our study (Young & Kindzierski 2023a). There are three instruments of interest with regard to validity of an IAT measurement of racial bias – the IAT itself (A); explicit measure(s) of bias, for example, indications of attitude, belief, or preference of bias captured in a questionnaire (B); and observations/measurements of real-world biased behaviors or actions (C). Rather clearly, real-world biased behaviors or actions (C) should rule. Thus, 'A' and 'B' should be positively correlated with 'C' for 'A' and 'B' to be taken as valid measures.

*2.1 Data Sets*

Oswald et al. (2013a) performed a meta-analysis of studies examining the predictive validity of IAT measures (A) and explicit measures (B) against measures of real-world behaviors (C) for a broad range of racial bias categories. These categories, used as proxies for racial and ethnic discrimination, included: brain activity, response time, microbehavior, interpersonal behavior, person perception, and policy/political preferences. Our interest was in data specific to racial discrimination specific to Black versus White groups (related to negative behaviors of Whites towards Blacks). Oswald et al. (2013a) research claims were:

- The IAT provides little insight into who will discriminate against whom, and provides no more insight than explicit measures of bias.
- Explicit measures of bias yielded predictions no worse than the IATs.

We intentionally used Oswald et al. (2013a) as our case study because of its publicly available data sets allowing us to employ p-value plots to confirm or refute their meta-analytic research claims. Our study only looked at two of their six categories of racial bias measures specific to Black versus White groups – microbehavior and person perception:

- They describe real-world microbehavior measures as… 'measures of nonverbal and subtle verbal behavior, such as displays of emotion and body posture during intergroup interactions and assessments of interaction quality based on reports of those interacting with the participant or coding of interactions by observers'.

- They describe real-world person perception measures as… 'explicit judgments about others, such as ratings of emotions displayed in the faces of minority or majority targets or ratings of academic ability'.

We extracted IAT and explicit bias meta-analytic data sets specific to Black versus White groups for the two categories from the Oswald et al. (2013a) supplemental files (Oswald et al. 2013b). These consisted of correlation coefficient ($r$) values and sample sizes ($n$) for the individual studies used in their meta-analysis for: 1) IAT−real-world behavior correlations, and 2) explicit bias−real-world behavior correlations.

Meta-analysis is used across multiple fields, including psychology. It is a procedure for combining test statistics from individual studies that examine a particular research question (Egger et al., 2001). A meta-analysis can evaluate a research question by taking a test statistic (e.g., a correlation coefficient between two variables of interest) along with a measure of its reliability (e.g., confidence interval) from multiple individual studies from the literature. The test statistics are combined to give, theoretically, a more reliable estimate of correlation between the two variables.

A requirement of meta-analysis is that test statistics taken from the individual studies for analysis are unbiased estimates (Boos & Stefanski, 2013). Given this requirement/assertion, independent evaluation of published meta-analysis on a particular research question has been used elsewhere to assess the statistical reproducibility of a claim coming from that field of research (Young & Kindzierski, 2019; Kindzierski et al., 2021; Young & Kindzierski, 2022, Young & Kindzierski, 2023b).

*2.2 Fisher's Z-transformation*

We first converted correlation coefficient (*r*) values to p-values using Fisher's test of a correlation method, also known as Fisher's *Z*-transformation (Fisher 1921, Wicklin 2017). Fisher's *Z*-transformation is a statistical technique used to compute a p-value for a correlation coefficient *r* given sample size *n*. The formula for Fisher's *Z*-transformation is:

$$Z = arctanh(r) = 0.5 \ln\left[\frac{(1+r)}{(1-r)}\right], \qquad [1]$$

which is considered to follow a normal distribution with standard error (*SE*) of:

$$SE = SQRT\left(\frac{1}{(n-3)}\right). \qquad [2]$$

A final step was to convert the average $\frac{Z-score}{SE}$ back into a p-value using the standard normal distribution.

*2.3 p-Value Plot*

The resulting p-values are presented in p-value plots. A p-value plot is used to visually check characteristics of a set of test statistics addressing the same research question. The plot – originally presented by Schweder & Spjøtvoll (1982) – is well-regarded, being cited more than 500 times in scientific literature (Google Scholar 2023).

In our case, the plot is used to assess whether the p-values follow a uniform distribution… if nothing is going on (i.e., there is little or no correlation) between two variables, the p-values for a data set should be evenly spread over the interval 0 to 1. The plot is a 2-way scatterplot where the observed p-values are ordered from smallest to largest and plotted against the integers: 1, 2, 3….n. If the observed data points fall approximately on a straight line in the plot, it suggests a good fit with the theoretical (uniform) distribution. In the absence of methodology and reporting biases (Ioannidis 2005 and 2022), and publication bias (Schimmack 2021), deviations from this straight line indicate departures from the uniform distribution and that there could be a real, non-random association.

## 3. Results

*3.1 Microbehavior measures*

3.1.1 Implicit bias (IAT)−real world behavior correlations

Here we examine correlations between IAT and negative microbehavior measures of Whites toward Blacks. A positive effect is an increase; whereas negative effects occur and can be taken as resulting from chance, no effect results. Table 1 of Oswald et al. (2013a) gives 87 correlations between IAT and microbehavior measures of Black versus White groups. Rank-ordered, the p-values computed for these 87 correlations are presented in Figure 1. Most of the p-values follow a 45-degree line in the plot indicating randomness.

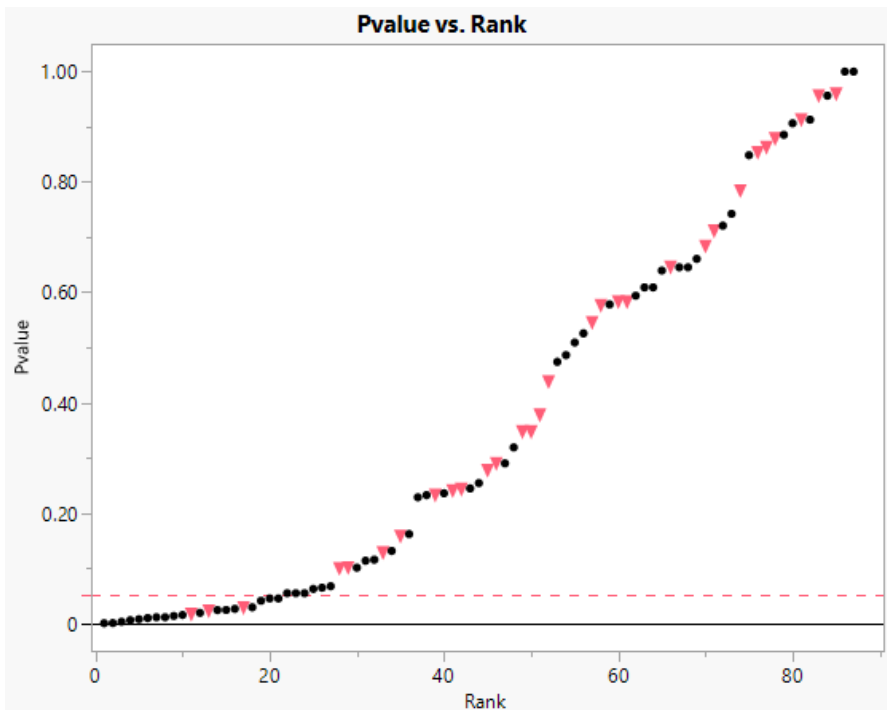

*Figure 1. P-value plot of 87 correlations between IAT results and real-world microbehaviors.*
Note: black circle (●) ≡ +ve correlation, i.e., IAT result is positively correlated with microbehavior; triangle (▼) ≡ −ve correlation, i.e., IAT result is negatively correlated with microbehavior.

There are multiple (30) negative correlations, i.e., an IAT result is negatively correlated with microbehavior, which are shown as downward pointing triangles (▼). Note that a downward pointing triangle, which according to psychology theory, should not be possible. These data points should be considered random (chance) results.

There are multiple (21) p-values less than 0.05 in Figure 1 including three decreases. A global test of the p-value distribution for the 87 p-values using Fisher's Combining of P-values (Fisher 1925, Young & Kindzierski 2021) gives a Chi-square of 322.51 with a p-value < 0.0001. The method assumes that p-values are independent of each other. However, this is not the case here as multiple p-values came from the same study (Oswald et al. 2013b).

To further evaluate the importance of small p-values in Figure 1, we computed multiplicity-adjusted p-values for the 10 smallest values using the method of Benjamini & Hochberg (1995). Unadjusted p-values and false discovery rate (FDR) adjusted p-values (after Benjamini & Hochberg 1995) for the 10 smallest p-values are listed in Table 1. Although some of the unadjusted p-values are small, the adjusted p-values are not impressive with only two smaller than 0.05 (see Table 1); and for these microbehaviors, most of the variability is due to other factors.

*Table 1. False Discovery Rate (FDR) p-values for 10 smallest unadjusted p-values in Figure 1.*

| Oswald et al. criterion description | Unadjusted p-value | FDR adjusted p-value |
|---|---|---|
| Cold | 0.000301 | 0.022733 |
| Speaking time | 0.000523 | 0.022733 |
| Hand/arm movement (load) | 0.002949 | 0.085534 |
| Speech errors | 0.005784 | 0.121068 |
| Expressive | 0.00766 | 0.121068 |
| Interactionally rigid | 0.009695 | 0.121068 |
| Smiling | 0.011133 | 0.121068 |
| Experiment's rating of interaction | 0.011133 | 0.121068 |
| Interactionally rigid | 0.013455 | 0.130064 |
| Seating selection | 0.015185 | 0.130617 |

In regard to their Oswald et al. (2013a) findings, elsewhere they stated (Oswald et al. 2015): ' …we found in our meta-analysis that across 87 effect sizes, the mean correlation between the race IAT and negative microbehaviors toward African Americans (sometimes called "microaggressions") was only .07, with a 95% confidence interval encompassing zero. This finding of a small and unreliable effect of IAT in the prediction of microbehaviors runs counter to common assertions that implicit bias primarily expresses itself through subtle negative behaviors during interracial interactions'.

Our independent p-value plot (Figure 1) agrees with Oswald et al.'s (2015) statements above. Our plot displays considerable randomness. Even if a few of the correlations might replicate, almost all or most of them are not expected to replicate. Our finding is that IAT evidence suggesting microbehaviors as a possible cause of racial disparities of Whites towards Blacks is unproven by the data and analysis.

3.1.2 Explicit bias−real world behavior correlations

Table 5 of Oswald et al. (2013a) gives 83 correlations between explicit bias and microbehavior measures. Rank ordered p-values computed for these 83 correlations are presented in Figure 2. As with Figure 1, most of the p-values follow a 45-degree line in the Figure 2 plot indicating randomness. There are multiple (33) negative correlations, i.e., explicit bias result is negatively correlated with microbehavior, which are shown as downward pointing triangles. Again, any decrease is assumed to be a random result.

We computed multiplicity-adjusted p-values for the 5 smallest values in Figure 2 using the method of Benjamini & Hochberg (1995). The unadjusted p-values and false discovery rate (FDR) adjusted p-values for the 5 smallest p-values are listed in Table 2. All 5 adjusted p-values are not impressive (see Table 2).

A global test of the p-value distribution for the 83 p-values using Fisher's Combining of P-values (Fisher 1925, Young & Kindzierski 2021) gives a Chi-square of 160.63 with a p-value = 0.603, indicating that nothing is going on. Keep in mind, the method assumes that the p-values are independent and that cannot be accepted here.

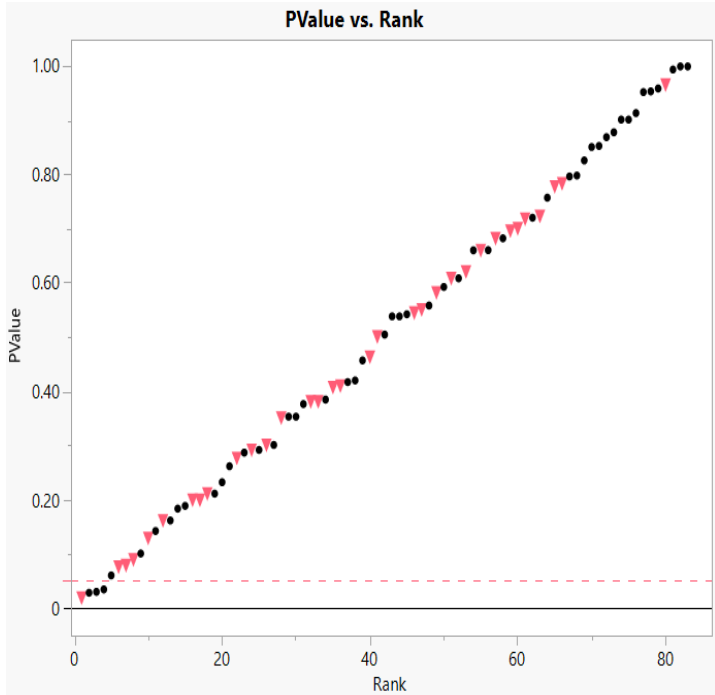

*Figure 2. P-value plot of 83 correlations between explicit bias measures and real-world microbehaviors.* Note: black circle (●) ≡ +ve correlation, i.e., explicit bias measure is positively correlated with microbehavior; triangle (▼) ≡ −ve correlation, i.e., explicit bias measure is negatively correlated with microbehavior.

*Table 2. False Discovery Rate (FDR) p-values for 5 smallest unadjusted p-values in Figure 2.*

| Oswald et al. criterion description | Unadjusted p-value | FDR adjusted p-value |
|---|---|---|
| Attitudes towards blacks, Carney | 0.0190 | 0.5762 |
| Attitudes towards blacks, Carney | 0.0284 | 0.5762 |
| Pro-black attitudes, Heider &… | 0.0301 | 0.5762 |
| Sem diff & F therm, McMonnell &… | 0.0346 | 0.5762 |
| Attitudes towards blacks, Carney | 0.0604 | 0.8631 |

Oswald et al. (2015) are silent about the (lack of) correlation between explicit bias and microbehaviors observed in their meta-analysis. Our p-value plot displays considerable randomness and although the 5 smallest unadjusted p-values are < 0.05, the adjusted p-values are > 0.05. Our p-value plot and the negative correlations are consistent with none of the explicit bias measure−real-world microbehavior correlations being real. Our finding is that even explicit bias measures do not support microbehaviors of Whites toward Blacks.

*3.2 Person perception measures*

3.2.1 Implicit bias (IAT)−real world behavior correlations

Correlations between IAT and negative person perception measures toward Blacks are examined here. As before, a positive effect is an increase; whereas negative effects occur and can be taken as chance or no effect results. Table 1 of Oswald et al. (2013a) gives 75 correlations between IAT and person perception measures.

Rank ordered p-values computed for these 75 correlations are presented in Figure 3. Most of the p-values follow a 45-degree line in the plot indicating randomness. There are multiple (26) negative correlations, i.e., IAT result is negatively correlated with person perception measures, which are shown as downward pointing triangles. Any decrease is assumed to be a chance (random) result.

A Fisher combined p-value (Fisher 1925, Young & Kindzierski 2021) was computed. The Chi-square value was 249.82 with a p-value < 0.0001. Multiplicity-adjusted p-values were computed for the 10 smallest values in Figure 3 using the method of Benjamini & Hochberg (1995). The unadjusted p-values and false discovery rate (FDR) adjusted p-values for the 10 smallest p-values are listed in Table 3. Although the 10 smallest unadjusted p-values are < 0.05, the adjusted p-values are not impressive with only one smaller than 0.05 (see Table 3). Again, the method assumes that p-values are independent of each other. This again is not the case here as multiple p-values came from the same study (Oswald et al. 2013b).

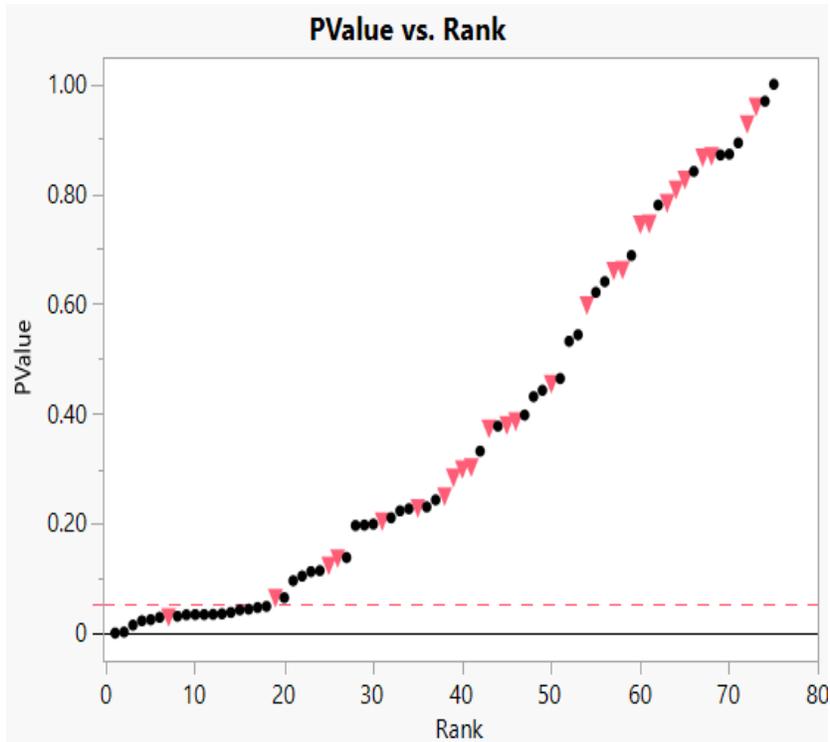

*Figure 3. P-value plot of 75 correlations between IAT results and real-world person perception measures.*
Note: black circle (●) ≡ +ve correlation, i.e., IAT result is positively correlated with microbehavior; triangle (▼) ≡ −ve correlation, i.e., IAT result is negatively correlated with microbehavior.

*Table 3. False Discovery Rate (FDR) p-values for 10 smallest unadjusted p-values in Figure 3.*

| Rank | R | N | Unadjusted p-value | FDR Adjusted p-value |
|------|------|-----|--------------------|----------------------|
| 1 | 0.582 | 31 | 0.000429 | 0.032196 |
| 2 | 0.4182 | 50 | 0.002256 | 0.084608 |
| 3 | 0.43 | 31 | 0.014952 | 0.204054 |
| 4 | 0.46 | 24 | 0.022669 | 0.204054 |
| 5 | 0.326 | 47 | 0.024811 | 0.204054 |
| 6 | 0.217 | 101 | 0.029044 | 0.204054 |
| 7 | -0.28 | 60 | 0.029859 | 0.204054 |
| 8 | 0.48 | 20 | 0.031059 | 0.204054 |
| 9 | 0.34 | 39 | 0.033624 | 0.204054 |
| 10 | 0.24 | 78 | 0.034022 | 0.204054 |

3.2.2 Explicit bias−real world behavior correlations

Table 5 of Oswald et al. (2013a) gives 79 correlations between explicit bias and person perception measures. Rank ordered p-values computed for these 79 correlations are presented in Figure 4. As with previous p-value plots, the p-values follow a 45-degree line in the Figure 4 plot indicating randomness. There are multiple (22) negative correlations, i.e., explicit bias result is negatively correlated with person perception, which are shown as downward pointing triangles. Again, any decrease is assumed to be a random (chance) result.

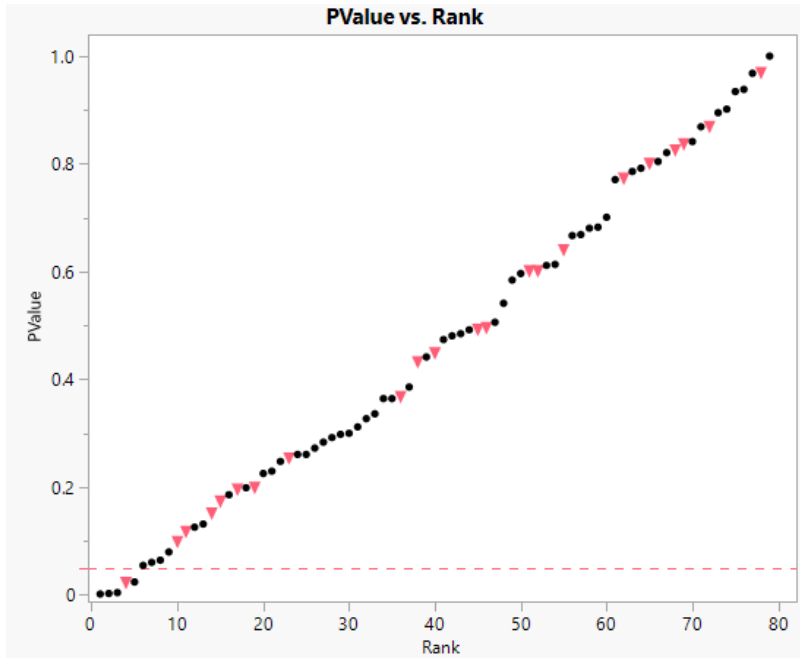

*Figure 4. P-value plot of 79 correlations between explicit bias measures and real-world person perception measures.*
Note: black circle (●) ≡ +ve correlation, i.e., explicit bias measure is positively correlated with microbehavior; triangle (▼) ≡ −ve correlation, i.e., explicit bias measure is negatively correlated with microbehavior.

A Fisher combined p-value (Fisher 1925, Young & Kindzierski 2021) was computed. The Chi-square value was 186.54 with a p-value of 0.0600. We computed multiplicity-adjusted p-values for the 5 smallest values in Figure 4 using the method of Benjamini & Hochberg (1995). The unadjusted p-values and false discovery rate (FDR) adjusted p-values for the 5 smallest p-values are listed in Table 4. All 5 adjusted p-values are not impressive (see Table 4).

*Table 4. False Discovery Rate (FDR) p-values for 5 smallest unadjusted p-values in Figure 4.*

| Rank | R | N | Unadjusted p-value | FDR Adjusted p-value |
|---|---|---|---|---|
| 1 | 0.700 | 16 | 0.001765 | 0.106167 |
| 2 | 0.506 | 32 | 0.002688 | 0.106167 |
| 3 | 0.320 | 77 | 0.004332 | 0.114069 |
| 4 | -0.383 | 35 | 0.022434 | 0.380790 |
| 5 | 0.390 | 33 | 0.024101 | 0.380790 |

## 4. Discussion

We considered three instruments of interest to the validity of an IAT measure of racial bias of Whites toward Blacks – Race the IAT itself (A); explicit measures of bias, for example, indications of attitude, belief, or preference of bias in a questionnaire (B); and observations/measurements of real-world biased behaviors or actions (C). The main issue is the validity of the IAT (measure of implicit bias) because of the popularity of the IAT and its widespread use in academic research and in business as a tool for dealing with DEI issues in the workplace. The IAT is a measurement that can be taken. Tested against zero (i.e., no comparison against anything) as a group measurement, it is statistically significant – not a chance result. However, many researchers – discussed earlier – do not consider the IAT to measure what it claims to measure, offering numerous explanations. Some of these are discussed further below.

Oswald et al. (2013a) point out that to be valid a measurement, the IAT must be confirmed by other measurements and be able to predict real-world results. Our p-value plots show that the IAT does not predict real-world microbehaviors (Figure 1) or person perception judgments (Figure 3). The IAT is a repeatable measure, but our figures support that it does not predict real-world behaviors, consistent with Oswald et al.'s (2013a) research claims. We add that the average correlation (*r*) of IAT reported by Oswald et al. (2013a) in their meta-analysis of IAT−microbehavior correlations specific to negative behaviors of Whites towards Blacks is small (0.07). The variance explained by a correlation is $r^2$; thus, the IAT explains less than one percent of the race variability, about the same as explicit bias. Ninety-nine percent of differences between the races is due to factors other than implicit bias.

Random behavior of p-values depicted in our plots also shows that explicit measures of bias poorly predict real-world microbehaviors (Figure 2) and person perception judgments (Figure 4) in the base studies used in the Oswald et al. (2013a) meta-analysis. Here Oswald et al. explains that the poor performance of both the IAT and measures of explicit bias are mostly consistent with a 'flawed instruments explanation' (i.e., problems in the theories that motivated development and use of these instruments). An explanation we agree with.

Getting back to the IAT, Schimmack (2019 and 2021) points out that the IAT is invalid as after psychometric (statistical) analysis; it is measuring essentially the same underlying construct as explicit measures of bias measure. Dang et al. (2020) note that the person-to-person variability of the IAT is small so that its ability to correlate with explicit or real-world measures of bias is very limited. Simplifying the argument… it is likely that the IAT is mostly measuring inherent human 'reaction time' to a stimulus (time it takes to detect, process, and respond to a stimulus), and this reaction time bears little or no relationship to subconscious thinking.

Additional literature observations support little or no explanatory power of race IAT (i.e., negative behaviors of Whites towards Blacks). First consider a systematic review by Dehon et al. (2017) examining evidence of the correlation between physician implicit racial bias based on the IAT and real-world, clinical decision making. Their systematic review considered nine studies (see Table 5)

Several comments of Table5 are in order. Most of the base studies are recent and eight of the nine have sufficient sample size to provide a well-powered test of the IAT versus zero. Table 5 essentially proves that the IAT is a very repeatable measure at a group level and greater than zero. The p-values versus zero are so small that we presented the $-Log10$ of the values. The largest p-value is $10^{-3.75}$ and the smallest is $10^{-189}$.

*Table 5. Dehon et al. (2017) systematic review results of studies examining the relationship between implicit racial bias based on the IAT and physician decision making.*

| 1st author | Year | n | Mean IAT score | SD | SE | -Log10 p-value | Rank | Assoc |
|---|---|---|---|---|---|---|---|---|
| Blair | 2014 | 138 | 0.30 | 0.29 | 0.025 | 32.44941 | 7 | ns |
| Cassell | 2015 | 216 | 0.40 | 0.43 | 0.029 | 42.55217 | 5 | ns |
| Green | 2007 | 287 | 0.36 | 0.40 | 0.021 | 65.14831 | 2 | s |
| Haider | 2014 | 248 | 0.41 | 0.48 | 0.024 | 64.70442 | 3 | ns |
| Haider | 2015 | 215 | 0.42 | 0.41 | 0.028 | 50.13419 | 4 | ns |
| Hirsh | 2015 | 129 | 0.50 | 0.42 | 0.037 | 40.88553 | 6 | ns |
| Oliver | 2014 | 543 | 0.43 | 0.34 | 0.0146 | 189.9263 | 1 | ns |
| Puumala | 2016 | 48 | | | | | | ns |
| Sabin | 2008 | 86 | 0.18 | 0.44 | 0.048 | 3.752433 | 8 | ns |

Note: n=sample size; SD=IAT standard deviation; SE=IAT standard error; Rank=p-value rank; Assoc=significant (s) or non-significant (ns) correlation between IAT and physician decision making.

Another way to examine whether a pattern/relationship exists in Dehon et al.'s results is to look at size of the standard deviations (SDs) relative to mean IAT scores. Large SDs relative to the means suggest that for all studies a substantial number of individuals had a negative IAT (no racial bias effect). Dehon et al. (2017) tested each IAT−clinical decision making (real-world action) correlation – refer to the far-right hand column of Table 5. They observed no significant correlation between the racial bias IAT and physician decision making in eight of nine studies – i.e., the racial bias IAT was not correlated with real-world physician decision making.

Robstad et al. (2019) recently examined Weight bias among 159 intensive care unit (ICU) nurses treating obese patients using the three instruments: the IAT (A), explicit bias questionnaire (B), and an anti-fat questionnaire of behavior intention (C). The latter – anti-fat questionnaire of behavior intentions – captured whether the ICU nurses followed acceptable treatment rules/algorithms using treatment vignettes.

Like previous IAT studies discussed here, Robstad et al. (2019) reported that, as a group, ICU nurses had a positive Weight IAT score suggesting a group bias against obese patients. However, the treatment vignettes showed that the nurses followed acceptable treatment guidance irrespective of the suggested Weight bias IAT score. In their analysis Robstad et al. noted that after explicit bias was considered, there was little or no predictive power of IAT in explaining behavior intentions of the nurses in their treatment of patients.

Robstad et al. (2019) computed correlation coefficient ($r$) values between implicit and explicit measures, age, work experience as ICU nurse, and an anti-fat questionnaire of behavioral intention. ICU nurses completed two IATs (Attitude and Stereotype) and their $r$ correlations with anti-fat questionnaire of behavior intentions were 0.11 and 0.03, respectively. Based on $r^2$, these would account for about 1% of the variance of their behavior intentions.

We extracted the Robstad et al. 78 correlations, converted them to p-values using Fisher's Z-transformation (Fisher 1921, Wicklin 2017), and constructed a p-value plot. This is shown in Figure 5. Although not shown here, there were many highly significant correlations (i.e., p-values << 0.05), yet the crucial correlations of the IATs (Attitude and Stereotype) with anti-fat behavior intention were not among these. Specifically, the Attitude IAT−behavior intention and Stereotype IAT−behavior intention p-values were 0.172 and 0.739, respectively.

Finally, Freichel et al. (2023) employed several machine learning methods to access the usefulness of implicit suicide cognitions, self-harm IATs, to predict concurrent desire to self-harm or die with an online community sample of 6,855 participants. Specifically, they assessed whether self-harm IATs add to the prediction capability of (explicit) concurrent self-reported suicidality (thoughts or ideas about the possibility of ending one's life) and desire to self-harm, over and above more easily collected measures – such as sociodemographic factors, self-reported history of self-harm and suicide, and explicit momentary self-harm and suicide cognitions.

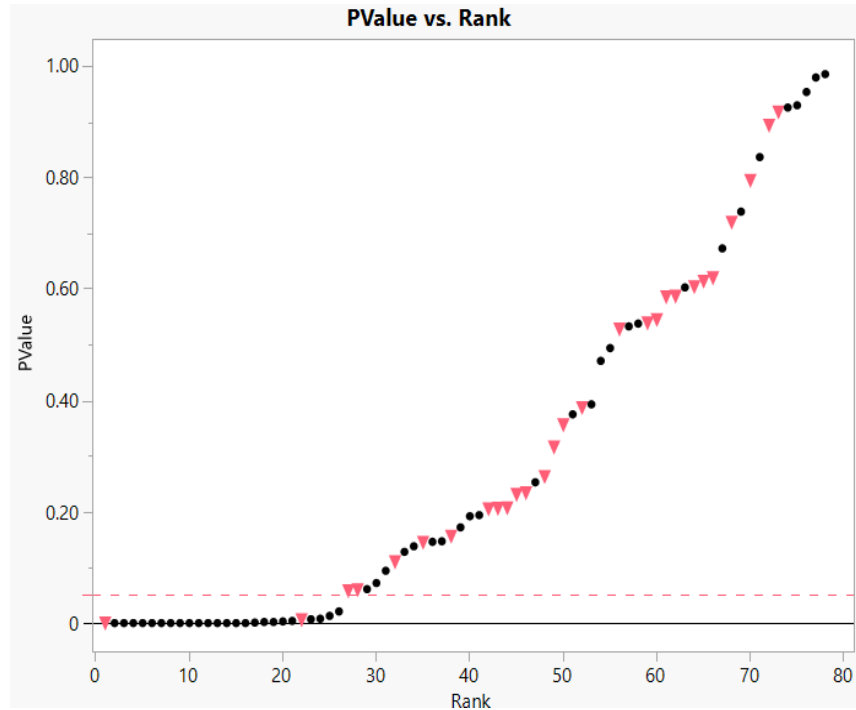

*Figure 5. P-value plot of 78 correlations between implicit and explicit measures, age, work experience as ICU nurse, and an anti-fat questionnaire of behavioral intention.*
Note: black circle (●) ≡ +ve correlation between two variables; triangle (▼) ≡ −ve correlation between two variables.

Freichel et al. (2023) observed that, in their best-performing model, self-harm, suicide, and death IATs offered very little (<2%) and often no predictive value on top of explicit measures that are much easier to collect to explain concurrent self-reported suicidality. Mood, explicit associations, and past suicidal thoughts and behaviors were the most important predictors of concurrent self-reported suicidality. Whereas implicit measures (self-harm, suicide, and death IATs) provided little to no gain in predictive accuracy. This study directly supports Shimmack's position that implicit bias and explicit bias measures are essentially the same; once explicit measures are considered, nothing is gained by using implicit measures.

Our p-value plots show that the IAT does not predict real-world negative microbehaviors (Figure 1) and person perception judgments (Figure 3) of Whites towards Blacks for the base studies used in the Oswald et al. (2013) meta-analysis. Our results are consistent with their findings. The plots also show that explicit bias measures do not predict real-world negative microbehaviors (Figure 2) and person perception judgments (Figure 3) of Whites towards Blacks. Again, our results are consistent with Oswald et al.'s (2013) findings.

Schimmack (2021) notes that over the past decade, it has become apparent that the empirical foundations of implicit social-cognition paradigm are problematic. He states that a key imposition that inhibits researchers within a paradigm from noticing these problems is publication bias. This bias ensures that studies that are consistent with a flawed paradigm (e.g., IAT is a valid assertion) are published and highlighted in review articles to offer false evidence supporting the paradigm. Indeed, once positive results are obtained, even if flawed, a research claim can become canonized (Nissen et al. 2016).

Does an IAT add any information over and above explicit measures of bias in explaining negative White against Black behaviors? Studies presented here support the claim that the IAT is an invalid metric when compared to real-world observations (i.e., it does not measure what it implies). So, the straightforward answer is that it does not add information over and above explicit measures of bias, which – as we observed here – itself also offers weak explanations of White against Black behaviors.

One of the most notable ideas of science is the use of experimentation (Fisher 1925): comparing treatment A to B and randomizing the placement of the treatments to experimental units. If the IAT is mostly a timing event measurement (i.e., measuring inherent human reaction time to a stimulus), there appear two fatal problems in IAT cases we have encountered in published literature outside the academic work of Oswald, Schimmack, Blanton, Tetlock, and others with similar findings:

- The timing event measurement is relative to zero. For example, we've not seen studies with direct comparisons of IAT measures for doctors versus lawyers or other highly-educated professionals (who may respond similarly because of their advanced educational training and comprehension skills).
- The timing event is given a name without establishing that the name is tied to the event (i.e., it is assumed that the event is a valid construct).

Until these can be addressed, current reproducible research supports that the White against Black (race) IAT is minimally or not at all correlated to explicit questionnaires or observed behaviors by people. Also, the question of whether the IAT measures are valid or a false direction for society remains unanswered.

## 5. Conclusions

Three instruments of interest were independently examined in a case study using statistical p-value plots regarding IAT measurement of Black−White racial bias – the IAT itself; explicit measure(s) of bias, for example, indications of attitude, belief, or preference of bias captured in a questionnaire; and observations/measurements of real-world biased behaviors or actions. The p-value plots exhibited considerable randomness for all IAT−real-world behavior and explicit bias−real-world behavior correlations examined. This randomness supports a lack of correlation between the IAT (implicit bias) and explicit bias measurements and real-world behaviors of Whites towards Blacks. These findings were for microbehaviors (measures of nonverbal and subtle verbal behavior) and person perception judgments (explicit judgments about others). Findings of the p-value plots were consistent with the case study research claim that the IAT provides little insight into who will discriminate against whom. It was also observed that the amount of real-world variance explained by the IAT and explicit bias measurements was small, less than 5%.